\documentclass[a4paper,12pt]{article}
\usepackage[utf8]{inputenc}
\usepackage{fontenc}
\usepackage{amsmath}
\usepackage{amsfonts}
\usepackage{amssymb}
\usepackage{amsthm}
\usepackage{commath}

\usepackage{caption}
\usepackage{multirow}

\usepackage{float}

\usepackage{graphicx}
\usepackage{cite}
\usepackage{verbatim}

\usepackage{color}

\usepackage[symbol]{footmisc}
\usepackage{titling}
\usepackage{sidecap}
\usepackage[total={6in, 9.5in}]{geometry}
\usepackage{hyperref}

 \hypersetup{
     colorlinks=true,
     linkcolor=blue,
     filecolor=magenta,      
     urlcolor=black,
     citecolor=red,
 }
\date{}
        
         \pdfoutput=1
\def\title#1{\begin{center} {\LARGE #1 \vspace{0.5cm}} \end{center}}
\def\author#1{\begin{center}{ \large #1} \end{center}}
\def\affil#1{\begin{center}{ \it #1} \end{center}}
\def\email#1{\begin{center}{\normalsize #1} \end{center}}
\def\preprint#1{\rightline{\begin{tabular}{l} #1 \end{tabular}}}
\def\acknowledgement#1{{\noindent \bf\large Acknowledgements} \,\,#1 }
\begin{document}
\preprint{} 

\title{Neutrino Oscillations and Non-standard \\[0.5cm] Interactions with KM3NeT-ORCA} 

\author{N. R. Khan Chowdhury on behalf of the KM3NeT Collaboration} 

\affil{IFIC - Instituto de Fisica Corpuscular (Univ. de Valencia - CSIC)}

\email{nafis.chowdhury@ific.uv.es}
\vspace{1cm}
%
%
\begin{abstract}

ORCA (Oscillations Research with Cosmics in the Abyss) is the low-energy node of KM3NeT, the next generation underwater Cherenkov neutrino detector in the Mediterranean sea. The primary goal of KM3NeT-ORCA is the determination of the neutrino mass ordering (NMO). With an energy threshold of few GeV and an effective mass of several Mtons, KM3NeT-ORCA can also make precision measurements of atmospheric oscillation parameters. Moreover, its access to a wide range of energies and baselines makes it optimal to discover exotic physics beyond the Standard Model such as Non-Standard Interactions (NSI) of neutrinos. The sensitivity of the detector to the neutrino mass ordering is presented, along with its potential for determination of the atmospheric oscillation parameters. It is observed that KM3NeT-ORCA will improve the current upper limits on NSI parameters by an order of magnitude after three years of data taking.
\end{abstract}
\vspace*{4cm}
\begin{center}
{\LARGE{Presented at}}\\
\vspace{1cm}
\Large{NuPhys2019: Prospects in Neutrino Physics\\
Cavendish Conference Centre, London, 16--18 December 2019}
\end{center}

\clearpage

 
\section{Introduction}
%
The conventional model of neutrino oscillations provides a successful interpretation of data taken by various solar \cite{Ahmad:2002jz}, atmospheric \cite{Fukuda:1998mi}, accelerator \cite{Abe:2011sj} and reactor \cite{Eguchi:2002dm, Abe:2011fz, An:2012eh} experiments. Crucial goals of future oscillation experiments can be grouped as (a) the determination of the neutrino mass ordering and the CP-violating phase $\delta$ with precise measurement of oscillation parameters, and (b) establishing the robustness of the standard three-flavor oscillation hypothesis with respect to physics beyond the Standard Model. In the present work we show the physics potential of KM3NeT-ORCA \cite{Adrian-Martinez:2016fdl} in trying to pin down these open questions.


The matter-induced modifications of neutrino oscillation probabilities is different for neutrino and anti-neutrino channels and is a function of neutrino mass ordering. In Fig. 1, if we consider the $P_{e \mu}$ and $P_{\bar{e} \bar{\mu}}$ appearance channels, one sees an enhancement in $P_{e\mu}$ and suppression in $P_{\bar{e} \bar{\mu}}$ if ordering is normal (NO) while if the ordering is inverted (IO), one gets the reverse. These ordering dependent alterations of oscillation signals discernible at few GeV give a handle to disentangle the two mass orderings.

\begin{figure}[ht]
   \centering	 	
    \includegraphics[width=0.49\linewidth]{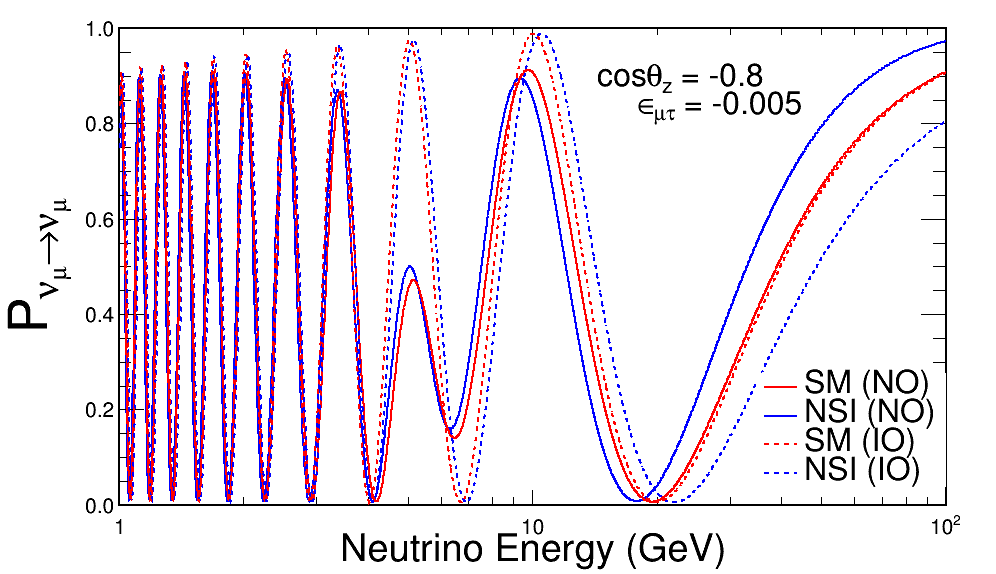}
    \includegraphics[width=0.49\linewidth]{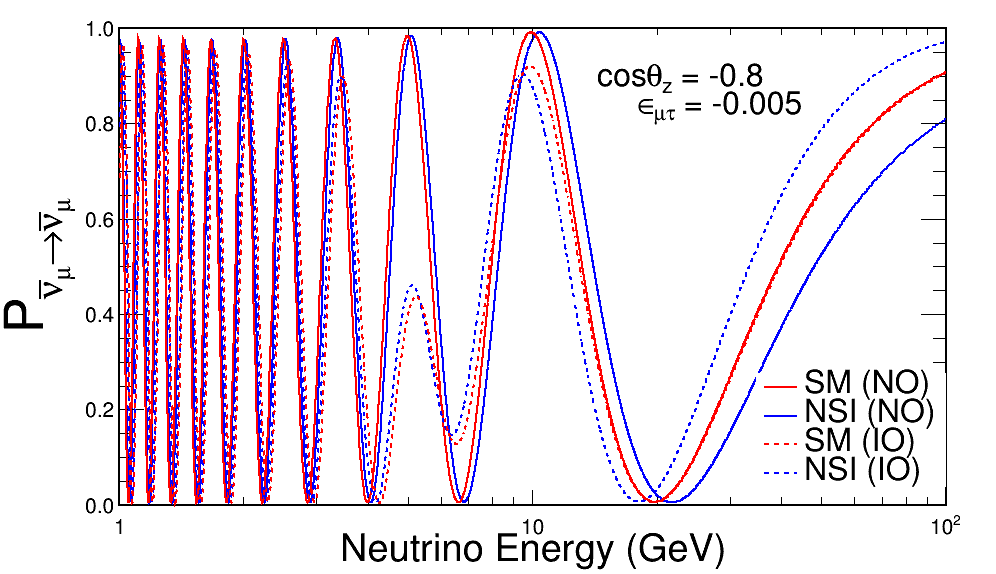}
    \includegraphics[width=0.49\linewidth]{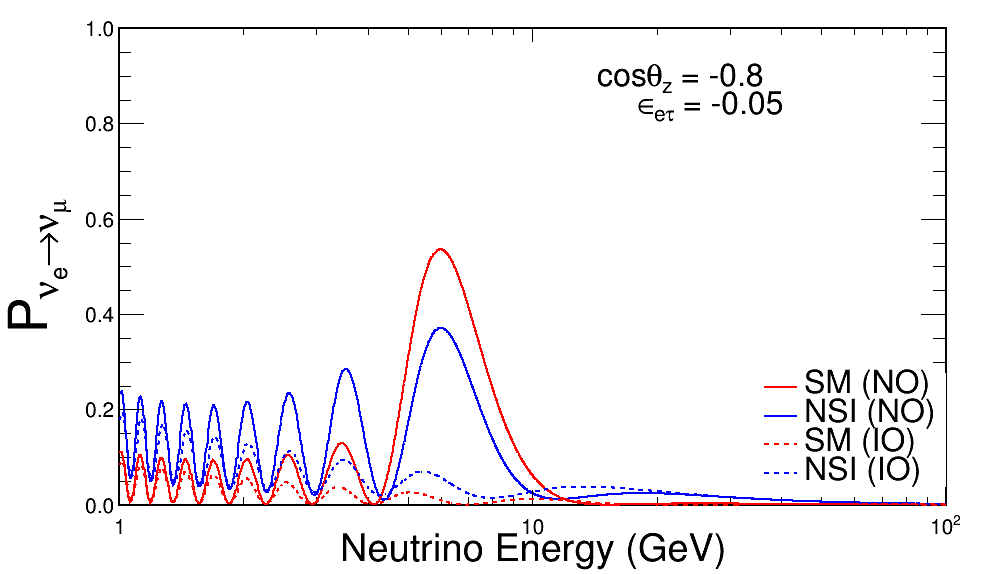}
    \includegraphics[width=0.49\linewidth]{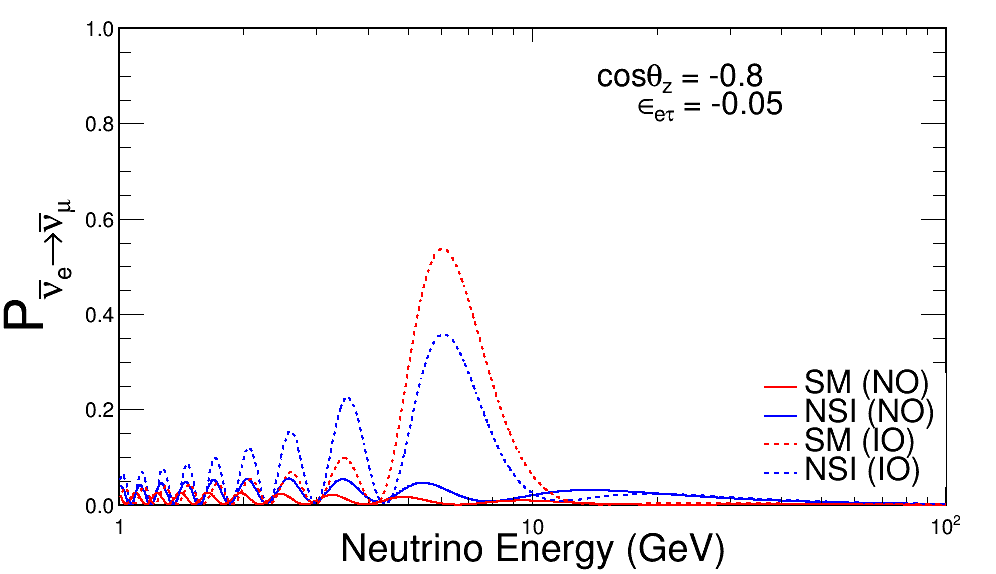}
   
\caption{Oscillation probabilities in the $\nu_{\mu}\rightarrow\nu_{\mu}$ (top) and $\nu_{e}\rightarrow\nu_{\mu}$ (bottom) channel as a function of neutrino energy for a fixed value of zenith angle ($\theta_{z}$). The solid (dashed) curves are for NO (IO). (Anti-)Neutrinos on the (right) left. The values of NSI parameters for which the blue curves are drawn are quoted. }
\end{figure}
In addition to the Standard Model (SM) MSW resonance, flavour changing neutral current
(NC) Non-Standard Interactions (NSI) \cite{Ohlsson:2012kf, Farzan:2017xzy} of neutrinos (of all flavours) with fermions ($e$, $u$ and $d$-quarks) present in Earth would alter the oscillation probabilities. NSI of neutrinos in propagation can be modelled as perturbations in the standard neutrino propagation Hamiltonian as,
\begin{equation}
H= \frac{1}{2E} U
\begin{bmatrix}
0&0&0\\
0&\Delta m^{2}_{21}&0\\
0&0&\Delta m^{2}_{31}
\end{bmatrix} U^{\dagger} 
+ 2\sqrt{2}G_{F}N_{f}(x) \begin{bmatrix}
1+ \epsilon_{ee}& \epsilon_{e\mu}& \epsilon_{e\tau}\\
\epsilon_{e\mu}^{\ast}&\epsilon_{\mu\mu}&\epsilon_{\mu\tau}\\
\epsilon_{e\tau}^{\ast}&\epsilon_{\mu\tau}^{\ast}&\epsilon_{\tau\tau}
\end{bmatrix}.
\end{equation}

\noindent
$G_{F}$ is the Fermi coupling constant, $N_{f}(x)$ is the fermion number density along the neutrino path, and the $\epsilon_{\alpha\beta}$ represent the NSI coupling parameters.  In this analysis, we consider non-standard interactions between neutrinos and $d$-quarks present in the Earth. The effect of the presence of non-standard interactions at one of the most dominant channels at ORCA is shown is Fig. 1. The effect for neutrinos in the Normal Ordering (NO)  is similar to anti-neutrinos in the Inverted Ordering (IO) .

\section{The KM3NeT-ORCA detector}
\begin{minipage}{0.6\linewidth}
The KM3NeT-ORCA detector \cite{Adrian-Martinez:2016fdl}, currently being installed at a depth of 2450 m in the Mediterranean Sea, is a megaton-scale water Cherenkov detector located 40 km offshore Toulon, France. Upon completion, the detector will consist of 115 detection units (DUs), each of which will comprise 18 spherical, 17" diameter Digital Optical Modules (DOMs) housing 31 3" PMTs and associated electronics. The average vertical spacing between the DOMs is 9 m and the horizontal spacing between the DUs is 23 m, amounting to a total instrumented volume of $\sim$8 Mton. The granularity of the detector layout makes it optimal to detect neutrinos with energies as low as $\sim$3 GeV.
 \end{minipage}%
\hspace{0.5cm} 
\begin{minipage}{0.4\linewidth}
\captionsetup{type=figure}
\includegraphics[width=0.96\linewidth]{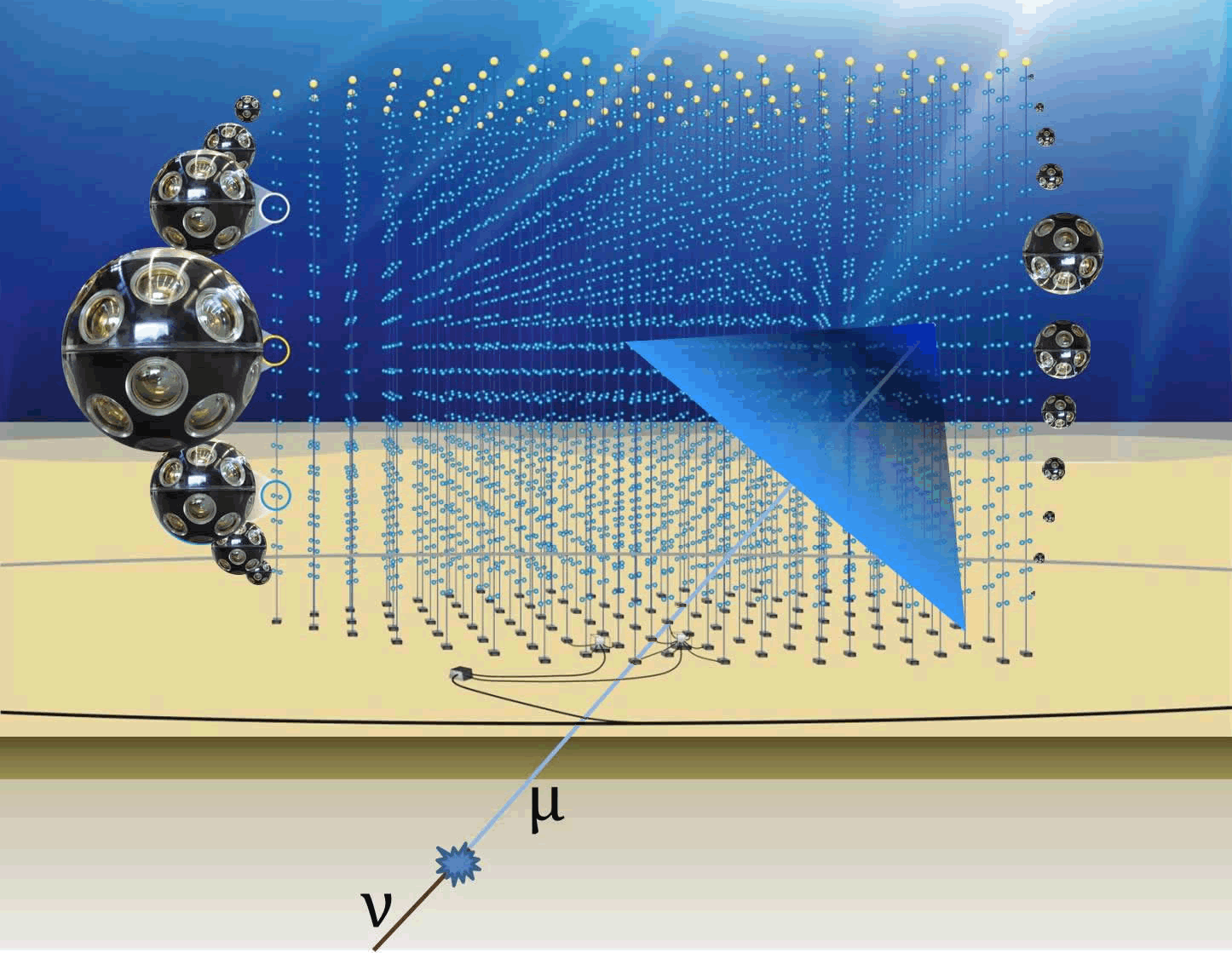}
\captionof{figure}{An artist's impression of the detector is shown.}
\end{minipage}

\section{Event spectra at the detector}
The HKKM 2014 \cite{Honda:2015fha} flux tables (Gran Sasso site)  are interpolated in $\log_{10}$(E) and cos$\theta_{z}$ and multiplied with the detector effective mass to calculate the rate of events for each interaction channel:  $\nu_{x}$ charged current (CC), $\bar{\nu}_{x}$ CC ($x$ = {e, $\mu$, $\tau$),  $\nu$ neutral current (NC) and $\bar{\nu}$ NC. Depending on the Cherenkov signatures of the outgoing lepton, two distinct event topologies are observed at the detector: track-like (left) and shower-like events (right). While $\nu_{\mu}$ CC interactions mostly account for track-like topology, shower-like topology has events from both $\nu_{e}$ CC and NC interactions.\\ 
  \begin{figure}[ht]
   \centering	 	
    \includegraphics[width=0.49\linewidth]{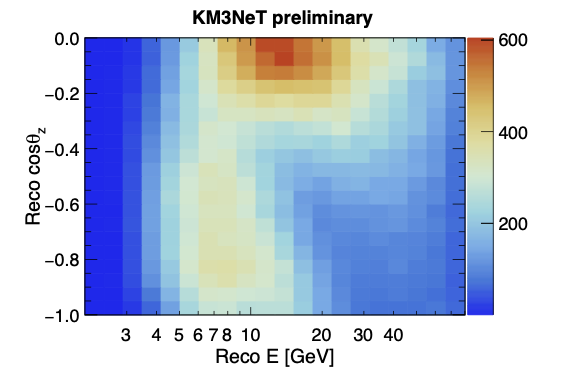}
 \includegraphics[width=0.49\linewidth]{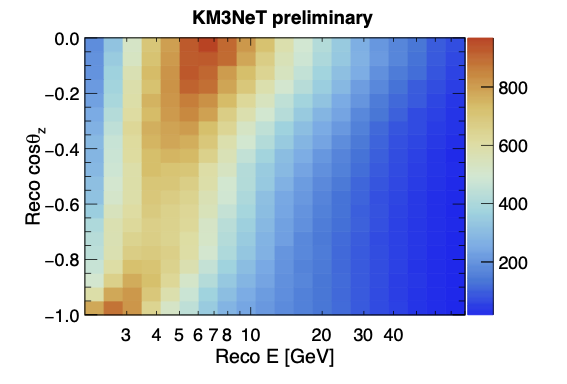}
	\caption{Two distinct event topologies at ORCA: tracks (left) and showers (right).
	Oscillation parameters are adopted from NuFit 3.2 \cite{Esteban:2016qun} with NO assumption.}
   \end{figure}

The statistical $\chi^{2}$ for each ($E,\theta_{z}$) bin is computed from,
\begin{equation}
\chi^{2}_{\rm E,\rm \theta_{z}}(\epsilon_{\alpha\beta}) = \frac{\Big(N^{\rm model}_{\rm E,\rm \theta_{z}}(\epsilon_{\alpha\beta}) -N^{\rm data}_{\rm E,\rm \theta_{z}}(\epsilon_{\alpha\beta}=0)\Big)\times\abs{N^{\rm model}_{\rm E,\rm \theta_{z}}(\epsilon_{\alpha\beta}) -N^{\rm data}_{\rm E,\rm \theta_{z}}(\epsilon_{\alpha\beta}=0)}}{N^{\rm data}_{\rm E,\rm \theta_{z}}(\epsilon_{\alpha\beta}=0) },
\end{equation}
where the superscript model represents the NSI case with $\epsilon_{e\tau} = -0.05$. All other NSI parameters are fixed at zero. Figure 4 shows the signed-$\chi^{2}$ maps for reconstructed events in the track-like  and
shower-like event topologies for three years of full KM3NeT-ORCA (115 DUs) runtime. 20 logarithmic bins were chosen in reconstructed neutrino energy ($E$) between 3 and 100 GeV, and 20 linear bins in cosine of the reconstructed zenith angle ($\theta_{z}$) between $-$1 and 0.

\begin{figure}[ht]
\includegraphics[width=0.47\textwidth,height=.35\textwidth]{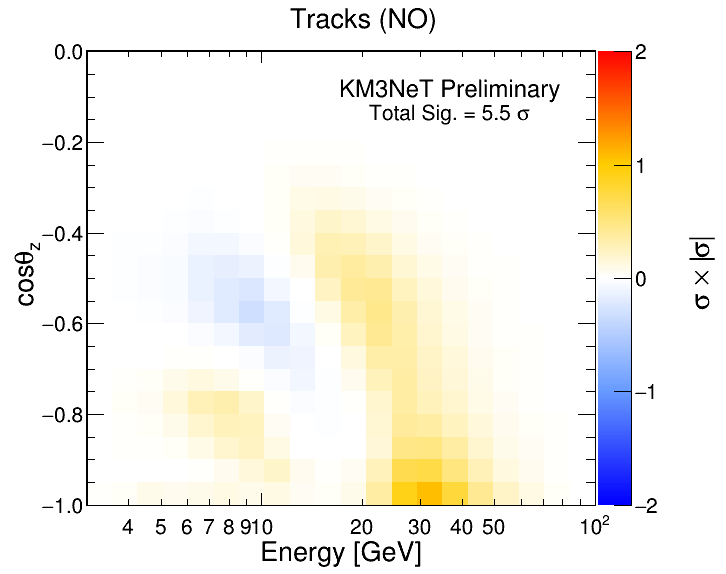}
\includegraphics[width=0.47\textwidth,height=.35\textwidth]{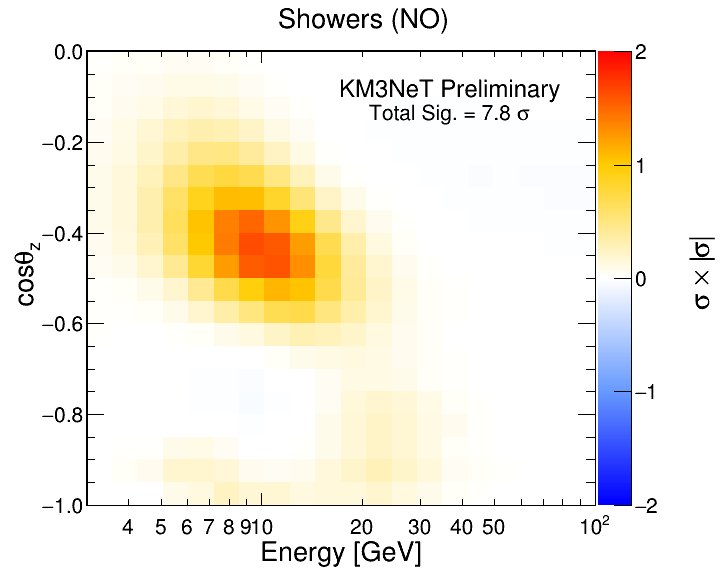}
\caption{Statistical $\chi^{2}$ as a function of reconstructed neutrino energy (E) and direction (cos$\theta_{z}$) for track-like (left) and shower-like (right) event topologies. NO is assumed. The colour code indicates the excess and deficits of events from the SM predictions. The total sensitivity quoted is the sum of the absolute value of the statistical $\chi^{2}$ for each bin.}
\end{figure}

\section{Systematics}

The final sensitivities of the experiment to NMO (or NSI) is estimated with a log-likelihood ratio test statistic based on the Asimov approach \cite{Gandhi:2007td}: 
\begin{multline}
 \chi^{2}_{{\rm NMO/NSI}} = 2\sum_{\rm E,\rm \theta_{z}} \Big (N^{\rm WO/NSI}_{\rm E,\rm \theta_{z}}(\epsilon_{\alpha\beta})\Big(1+\sum_{k}f^{k}_{\rm E,\rm \theta_{z}}\zeta_{k}\Big) -N^{\rm RO/SM}_{\rm E,\rm \theta_{z}}(\epsilon_{\alpha\beta}=0)\\
 +  N^{\rm RO/SM}_{\rm E,\rm \theta_{z}}(\epsilon_{\alpha\beta}=0) \ln\frac{N^{\rm RO/SM}_{\rm E,\rm \theta_{z}}(\epsilon_{\alpha\beta}=0)}{N^{\rm WO/NSI}_{\rm E,\rm \theta_{z}}(\epsilon_{\alpha\beta})\Big(1+\sum\limits_{k}f^{k}_{\rm E,\rm \theta_{z}}\zeta_{k}\Big)}\Big ) + \sum_{k}\zeta_{k}^{2}.   
\end{multline}

$N^{\rm RO}_{\rm E,\theta_{z}}$ ($N^{\rm WO}_{\rm E,\theta_{z}}$) denotes the expected number of track / shower events in a given $[\rm E, \rm \theta_{z}]$ bin for the right ordering (wrong ordering) hypothesis in the standard 3$\nu$ oscillation framework. In the case of NSI sensitivity estimation, $N^{\rm NSI}_{\rm E,\rm \theta_{z}}$ ($N^{\rm SM}_{\rm E,\rm \theta_{z}}$) is the predicted number of events for an assumed mass ordering in presence (absence) of NSI. \\

\begin{minipage}{0.38\linewidth}
Systematics are included in our simulation using the ``pull'' method \cite{Gandhi:2007td, Ghosh:2012px}. Table 1 lists nuisance parameters and oscillation parameters adopted from NuFit 3.2 \cite{Esteban:2016qun} and their corresponding Gaussian priors (if any) over which marginalisation has been done to minimise the value of $ \chi^{2}_{{\rm NMO/NSI}}$ . 
The individual contributions from track-like and shower-like events are added in quadrature to compute the total significance. 

\end{minipage}
\hfill
\hspace{0.5cm}
\begin{minipage}{0.6\linewidth}
\begin{center}
\captionsetup{type=table}
\captionof{table}{List of systematics.} \label{tab:title} 
 \begin{tabular}{||l c c c||} 
  \hline
 \textbf{parameters}&\bf{treatment}&\textbf{true values} &\textbf{prior}   \\

\hline
$\Delta m^{2}_{21}/10^{-5} eV^{2}$&fix&7.40   & - \\
$\Delta m^{2}_{31}/10^{-3} eV^{2}$&fitted& 2.494 & free \\
$\theta_{12}(^\circ)$ &fix&33.62 & -\\
$\theta_{13} (^\circ)$ &fitted& 8.54 & 0.15\\
$\theta_{23} (^\circ)$ &fitted&47.2& free\\
$\delta_{CP} (^\circ)$ &fitted&234&  free\\
Flux norm. & fitted&1 & 10$\%$ \\
NC scale &fitted& 1 & 5$\%$\\
Energy slope &fitted& 1 & 3$\%$\\
$\nu_{\mu}/\nu_{e}$  skew &fitted& 0 & 5$\%$\\
$\nu/\bar{\nu}$ skew &fitted& 0 & 3$\%$\\
\hline

\end{tabular}
\end{center}
\end{minipage}
\hspace{0.5cm}

\section{Results}
\subsection{Neutrino Oscillations}

The Asimov NMO sensitivity \cite{PoS(ICRC2019)1019} for three years of KM3NeT-ORCA runtime is shown in Fig. 5 (left) for a range of possible true $\theta_{23}$ values . The curves are drawn for both assumed true orderings and the most favourable and least favourable $\delta_{CP}$ values.

\begin{figure}[H]
\centering
\includegraphics[width=.49\linewidth]{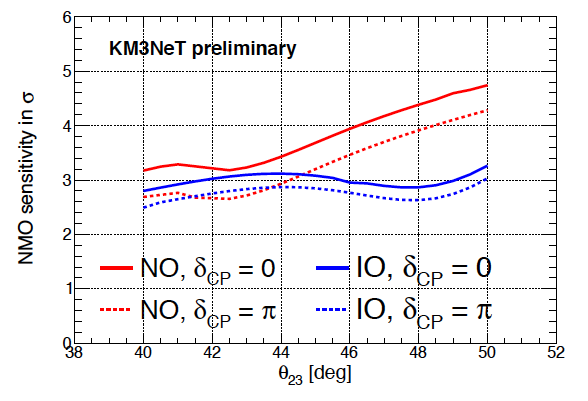}
\includegraphics[width=.49\linewidth]{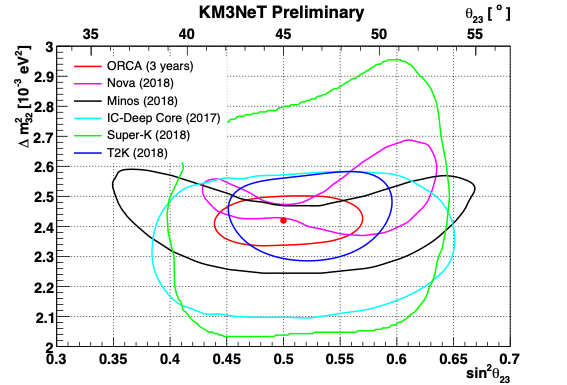}
\caption{Projected sensitivity to NMO (left) for truth NO (red) and IO (blue) assumptions.
Exclusion plot in $\theta_{23}$ - $\Delta m^{2}_{32}$ plane on the right .}
\end{figure}

Allowed region of atmospheric oscillation parameters by KM3NeT-ORCA \cite{PoS(ICRC2019)1019} after three years of running is shown in Fig. 5 (right),
overlapped with current constraints from MINOS \cite{Habig:2010vw}, NO$\nu$A \cite{Acero:2019ksn}, Super-K \cite{Ashie:2005ik} and IceCube-DeepCore \cite{Aartsen:2017nmd}. The 90\% CL contour is drawn assuming NO (fixed) and $\delta_{CP}$ = 0 (fitted).


\subsection{Non-standard Interactions}
The 90\% C.L. contours in correlated NSI parameter spaces allowed after three years of data taking of KM3NeT-ORCA are shown for both orderings assumptions. The NSI parameters not appearing on the plots are fixed at zero.

\begin{figure}[ht]

\includegraphics[width=0.48\textwidth,height=.35\textwidth]{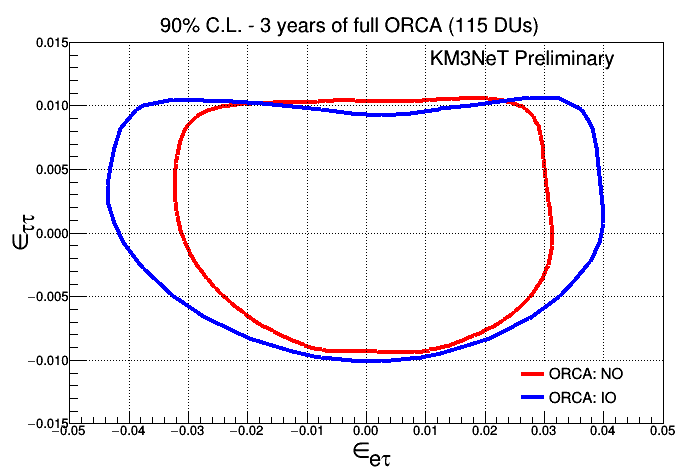}
\includegraphics[width=0.48\textwidth,height=.35\textwidth]{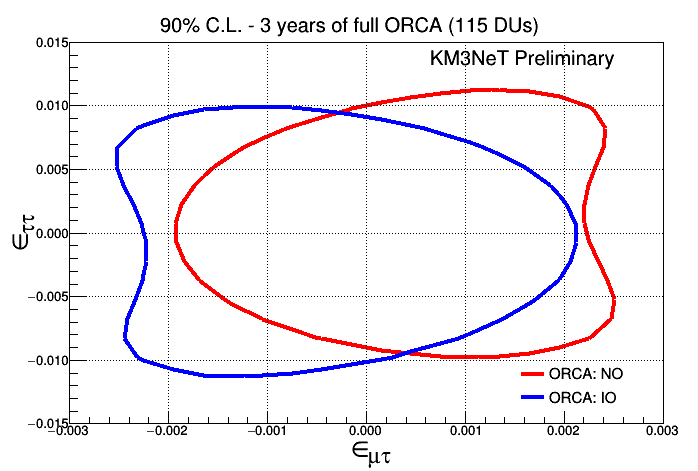}

\caption{Allowed region in correlated NSI parameter phase spaces $|\epsilon_{e\tau} -\epsilon_{\tau\tau}|$ (left) and $|\epsilon_{\mu\tau} -\epsilon_{\tau\tau}|$ (right) after three years of ORCA runtime. Pseudo data is generated for ($\epsilon_{ij}$, $\epsilon_{kl}$) = (0, 0).}
\end{figure}


\begin{figure}[ht]
\centering

\includegraphics[width=.49\linewidth]{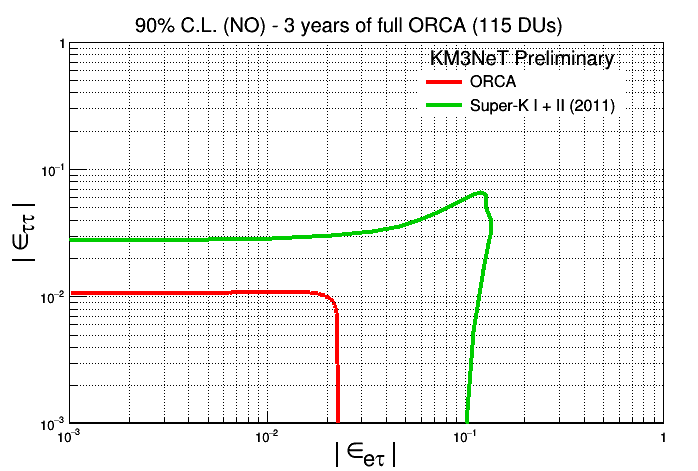}
\includegraphics[width=.49\linewidth]{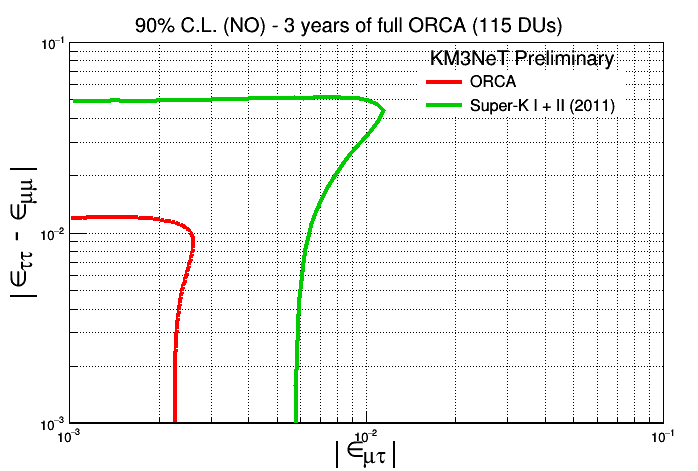}
			\caption{Allowed NSI parameter regions in the $e - \tau$ sector (left) for $\epsilon_{ee}$ = 0  and  $\mu - \tau$ sector (right) in the \textit{hybrid model} approximation \cite{Mitsuka:2011ty} are shown. }
\end{figure}

\begin{minipage}{0.42\linewidth}
\captionsetup{type=figure}
\includegraphics[width=.95\linewidth]{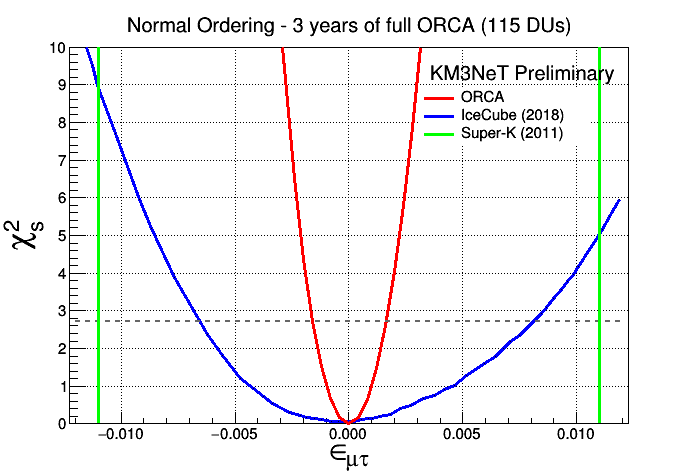}	
\captionof{figure}{Sensitivity to $\epsilon_{\mu\tau}$.}

\end{minipage}%
\begin{minipage}{0.56\linewidth}
In Figure 7 and 8, the exclusion region assuming NO in the \textit{hybrid model} approximation ($\theta_{12},\theta_{13}$, and $\Delta m^{2}_{21}=0$)  is drawn for comparison with IceCube \cite{Aartsen:2017xtt} and Super-K \cite{Mitsuka:2011ty}. With three years of run time, ORCA is expected to constrain NSI parameters $\epsilon_{e\tau}$, $\epsilon_{\mu\tau}$ and $\epsilon_{\tau\tau}$ by a factor of four better than current limits.
\end{minipage}


\section{Summary}
In this contribution, future projections of sensitivity of KM3NeT-ORCA towards neutrino mass ordering resolution and precise measurement of atmospheric oscillation parameters has been reported. The impact of non-standard interactions on the event signal at KM3NeT-ORCA is probed and its discovery potential to different NSI phase spaces has been discussed. 
\newline

\acknowledgement{We gratefully acknowledge the financial support of the Ministry of Science, Innovation and Universities: State Program of Generation of Knowledge, ref. PGC2018-096663-B-C41 (MCIU / FEDER), Spain. }

\bibliographystyle{unsrt}
\bibliography{ms.bib}

\end{document}